\documentclass[aps,prl,twocolumn,superscriptaddress]{revtex4-1}

\usepackage{graphicx}
\usepackage{tabularx}
\newcolumntype{Y}{>{\centering\arraybackslash}X}
\usepackage{bm}
\usepackage{color}
\usepackage{amsmath}

\begin{document}

\def\mos2{$\mathrm{Mo S_2}$}
\def\mose2{$\mathrm{Mo Se_2}$}
\def\ws2{$\mathrm{W S_2}$}
\def\wse2{$\mathrm{W Se_2}$}

\title{First principles coupled cluster theory of the electronic spectrum of the transition metal dichalcogenides}

\author{Artem Pulkin}
\email{Present address: QuTech and Kavli Institute of Nanoscience, Delft University of Technology, 2600 GA Delft, The Netherlands \\
E-mail: gpulkin@gmail.com}
\author{Garnet Kin-Lic Chan}
\email{E-mail: gkc1000@gmail.com}
\affiliation{Division of Chemistry and Chemical Engineering, California Institute of Technology, Pasadena, CA 91125, USA}

\date{\today}

\def\artem#1{\textcolor{red}{#1}}

\begin{abstract}
  The electronic properties of two-dimensional transition metal dichalcogenides (2D TMDs) have attracted much attention during the last decade.
  We show how a diagrammatic \textit{ab initio} coupled cluster singles and doubles (CCSD) treatment
  paired with a careful thermodynamic limit extrapolation in two dimensions can be used to
  obtain converged bandgaps for monolayer materials in the \mos2 family.
  We find general agreement between CCSD and previously reported $GW$ simulations in terms of the band structure,
  but predict slightly higher band gap values and effective hole masses compared to previous reports.
  We also investigate the ability of CCSD to describe  trion states, finding reasonable qualitative structure,
  but poor excitation energies due to the lack of screening of three-particle excitations in the effective Hamiltonian.
  Our study provides an independent high-level benchmark of the role of many-body effects in 2D TMDs and showcases
  the potential strengths and weaknesses of diagrammatic coupled cluster approaches for realistic materials.
\end{abstract}

\pacs{}

\maketitle

\noindent

Two-dimensional (2D) materials are flat, atomically thin crystals\cite{novoselov_two-dimensional_2005, butler_progress_2013, kaul_two-dimensional_2014}
that can be prepared through a number of well-established techniques\cite{splendiani_emerging_2010, lee_synthesis_2012, zhang_direct_2014}.
They display diverse electronic structure ranging from conventional semiconductors with an optical band gap\cite{mak_atomically_2010}
to topological insulators\cite{kane_quantum_2005, qian_quantum_2014},
materials hosting unconventional magnetism\cite{samarth_condensed-matter_2017, gibertini_magnetic_2019},
superconductivity\cite{fatemi_electrically_2018, xu_topological_2018}
and spin physics\cite{xiao_coupled_2012, xi_ising_2016}.
2D transition metal dichalcogenides (TMDs) $MX_2$, $M = \mathrm{Mo,W,Nb,Ti,...}$, $X = \mathrm{S,Se,Te}$ are amongst the most
promising representatives of this family, with potential applications in electronics\cite{splendiani_emerging_2010, radisavljevic_integrated_2011, radisavljevic_single-layer_2011, wang_electronics_2012, baugher_intrinsic_2013, duan_two-dimensional_2015}
and beyond\cite{xiao_coupled_2012, qian_quantum_2014, fatemi_electrically_2018}.

The electronic properties of the 2H structural phase of
monolayer TMDs and \mos2 in particular have been studied extensively through spectroscopy\cite{mak_control_2012, bernardi_extraordinary_2013, ugeda_giant_2014, wang_electronics_2012, ugeda_observation_2018},
and transport\cite{radisavljevic_single-layer_2011, wang_electronics_2012} experiments.
Several studies have probed the electronic band structure directly~\cite{mak_atomically_2010, splendiani_emerging_2010, mak_control_2012, mak_tightly_2013, jin_direct_2013, alidoust_observation_2014, jin_substrate_2015, vancso_intrinsic_2016, hill_band_2016}.
A large number of theoretical works based on model or first-principles Hamiltonians (the latter in conjunction with
various flavors of density functional theory (DFT) and the DFT-$GW$ approximation)
have been carried out to complement the experimental picture~\cite{
    li_electronic_2007,
    zhu_giant_2011,
    ellis_indirect_2011,
    cheiwchanchamnangij_quasiparticle_2012,
    johari_tuning_2012,
    kumar_electronic_2012,
    ramasubramaniam_large_2012,
    qiu_optical_2013,
    shi_quasiparticle_2013,
    zahid_generic_2013,
    qiu_screening_2016}.
Although broad agreement between theory and experiment is observed, there is uncertainty at the quantitative
level in numerical results arising from different methodologies, with estimates of the bandgap varying over a range of 0.4 eV or more
even from different flavours of DFT-$GW$~\cite{qiu_screening_2016}.
In this context, independent high-level many-body benchmarks of the electronic properties of monolayer TMDs remain desirable.

   The $GW$ approximation  has long been the state-of-the-art in the \textit{ab initio} many-body description
   of materials  single-particle spectra. More recently, however, coupled cluster (CC) theory, widely
   used in accurate molecular calculations, has emerged as a viable approach for extended systems~\cite{hirata_highly_2001, hirata_coupled-cluster_2004, katagiri_equation--motion_2005, booth2013towards,yang2014ab, mcclain_gaussian-based_2017}. In particular,
   equation-of-motion coupled cluster theory, the extension to excited states and spectra, has been
   shown at the singles and doubles level of approximation (EOM-CCSD) to yield improved results over the GW approximation in more
   correlated systems, with a reduced    dependence on the mean-field starting point~\cite{mcclain2016spectral}, while a careful diagrammatic analysis
   shows that EOM-CCSD is diagrammatically comparable or superior to the standard $GW$ approximation\cite{lange_relation_2018}.

   In the present work we report benchmark single-particle bandgaps for the monolayer TMDs and the full band structure for the
 representative monolayer material \mos2, obtained
 from first-principles  EOM-CCSD calculations. 
We discuss the role of dimensionality in finite size effects which are the largest source of uncertainty in our results.
The bandgaps and the \mos2 band structure  are generally in good agreement with previous many-body results,
and we analyze remaining uncertainties in our computed band structure arising from spin-orbit coupling.
Finally, we examine the three-particle excited states of the coupled cluster Hamiltonian and compare them to trions observed in these materials.

We describe the 2D material in terms of an all-electron Hamiltonian with periodic boundary conditions (PBC) expressed in a finite crystalline
molecular orbital (Bloch) basis: 
\begin{equation}
    H = 
    \sum_{\alpha \beta k} h_{\alpha \beta k}
    \mathbf{c}^\dagger_{\alpha k}
    \mathbf{c}_{\beta k} + 
    \sum_{\substack{\alpha \beta \gamma \delta \\ k_1 k_2 k_3}} v_{\substack{(\alpha k_1), (\beta k_2) \\ (\gamma k_3), (\delta k_4)}}
    \mathbf{c}^\dagger_{\alpha k_1}
    \mathbf{c}^\dagger_{\beta k_2}
    \mathbf{c}_{\delta k_4}
    \mathbf{c}_{\gamma k_3} ~,
    \label{eq:h}
\end{equation}
where $\mathbf{c}_{\alpha k}$ is a fermionic annihilation operator removing one electron from crystalline molecular spin-orbital $\alpha k$ ($\alpha$ and other Greek symbols denote orbitals, $k$ is the Bloch wavevector),
$h$ denotes single-particle Hamiltonian matrix elements corresponding to the sum of kinetic and external potential energies of the electrons,
$v$ denotes the Coulomb repulsion matrix element for the four crystalline molecular orbitals ($k_4$ satisfies pseudomomentum conservation in the scattering process: $k_1 + k_2 = k_3 + k_4$).
The crystalline molecular orbitals $\psi_{\alpha k} (r)$ are  expanded in terms of crystalline Gaussian atomic orbitals $\tilde{\phi}_{\lambda k}$, i.e. $\psi_{\alpha k} = \sum_\lambda u_{\alpha\lambda} \tilde{\phi}_{\lambda k }$, with
$\tilde{\phi}_{\lambda k} (r) = \sum_R \phi_\lambda (r - R) e^{i k \cdot R}$ where
$\phi_\lambda (r)$ is a Gaussian basis function centered at $r=0$; $R$ are lattice vectors.
In this work we used the DZVP basis\cite{godbout_optimization_1992} for dense $k$-point calculations of monolayer \mos2, \mose2 and the def2-TZVP\cite{weigend_balanced_2005} basis set was used for the tungsten materials.
The wavevectors $k$ were chosen to sample the Brillouin zone (BZ) uniformly.
The Coulomb integrals $v$ can  formally be defined to be finite in reciprocal space:
\begin{multline}
    v_{\substack{(\alpha k_1), (\beta k_2) \\ (\gamma k_3), (\delta k_4)}} = v_{1234} = \int d G \cdot K(G) \rho_{13} (G) \rho_{24} (-G) \approx \\
    \sum_{G \neq 0} w \cdot K(G) \rho_{13} (G) \rho_{24} (-G) ~,
    \label{eq:v}
\end{multline}
where $G$ are reciprocal lattice vectors, $K(G) = \int dr \cdot e^{-i G \cdot r} / r$ is the Coulomb kernel; $\rho_{ij} (G) = \int dr e^{-i G \cdot r} \psi_i^* (r) \psi_j (r)$ are Fourier transforms of pair densities ($r$ belongs to the unit cell);
$w$ is the volume of the reciprocal-space grid box. In this work, we used Gaussian density fitting~\cite{sun_python-based_2017,sun2017gaussian}
to compute $v$ efficiently.

\begin{figure}
    \includegraphics{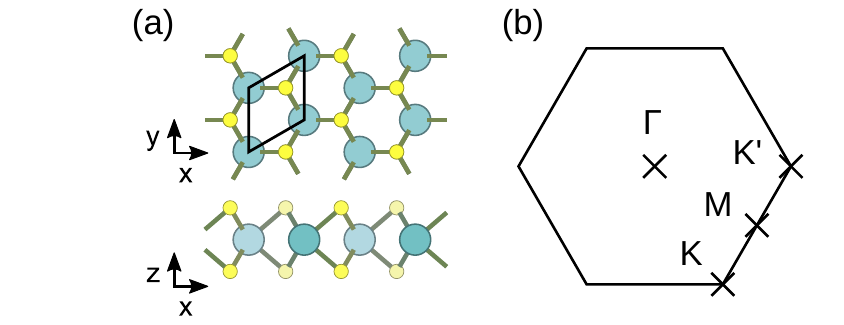}
    \includegraphics{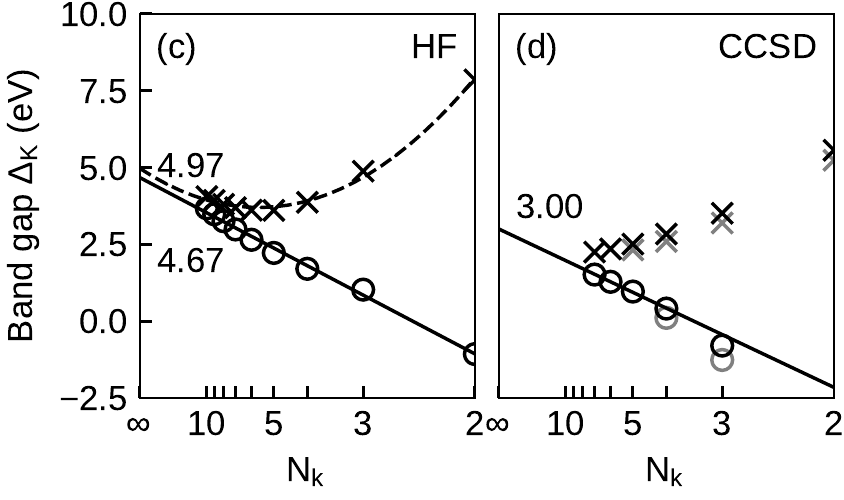}
    \caption{
       Atomic structure and the size of the Hartree-Fock (HF) and coupled-cluster singles-and-doubles (CCSD) band gap in monolayer \mos2.
        (a) Atomic structure of monolayer \mos2: top and side views.
        Sage circles indicate the location of Mo atoms, yellow circles stand for S atoms.
        (b) The location of high-symmetry points $\Gamma$, M, K, K' in the hexagonal Brillouin zone.
        (c,d) Extrapolation of HF and CCSD band gaps at the high-symmetry K point $\Delta_\mathrm{K}$.
        Solid lines and circles are obtained from the 2D approach when calculating the band gap; dashed line and crosses correspond to the 3D approach
        (see main text).
        The CCSD band gaps calculated using the full DZVP basis set are presented as a reference (grey symbols).
        The horizontal axis represents the number of $k$-points along one reciprocal axis plotted on the $N_k ^ {-1}$ scale.
        Scatter and line plots represent calculated and fitted values of the band gap respectively.
        The extrapolated $N_k \rightarrow \infty$ band gap sizes are indicated.
        Negative values of the band gap reflect the overlap of occupied and virtual single-particle spaces during the HF calculation due to the finite-size error discussed in the main text.
    }
    \label{fig:1}
\end{figure}

We constructed the Hamiltonian in Eq.~\ref{eq:h} and determined the spectrum in the following steps:
\begin{enumerate}
\item The Hartree-Fock determinant $|\Phi\rangle$ and molecular orbitals were computed in the crystalline Gaussian orbital basis.
  Real-space summation cutoffs and other saturating parameters where chosen to produce sub-$\mu$Hartree error in the total energy in
  a $3 \times 3$ $k$-point  benchmark model;
      \item An ``active'' single-particle space for the correlated calculations was  restricted to the 7 hole and 7 electron bands adjacent to the band gap, see Fig.~\ref{fig:1}(d) for the basis set error in the resulting band gap size;
    \item The Hamiltonian matrix elements were re-computed within this active space;
    \item The ground-state wavefunction $e^T|\Phi\rangle$ was computed at the (spin-restricted) coupled-cluster singles and doubles (CCSD) level\cite{mcclain_gaussian-based_2017}, where $T$ contains one and two particle-hole excitations;
    \item Low-lying equation-of-motion (EOM) spin-restricted CCSD roots were determined in the $N_e+1$ (electron affinity, EA) and $N_e-1$ (ionization potential, IP) sectors ($N_e$ is the number of electrons per unit cell multiplied by the number of $k$ points sampled in the Brillouin zone) by diagonalizing
      the coupled cluster effective Hamiltonian $\bar{H}=e^{-T}He^T$ in the p, pph (EA) and h, phh (IP) spaces (where p and h denote particle, hole
      respectively)~\cite{stanton1993equation,krylov2008equation,shavitt2009many,mcclain_gaussian-based_2017}.
        This was done for a set of uniformly shifted $k$ point grids, where shift vectors sample the Brillouin zone of the supercell uniformly: for example, values on an $18 \times 18$ $k$ point grid were obtained through $6$ separate $6 \times 6$ ground- and excited-state CCSD calculations (time reversal symmetry was used).
The principal roots were interpreted as the band energies.
\end{enumerate}
All calculations were performed using the \texttt{PySCF} package\cite{sun_pyscf:_2018}.

Apart from the intrinsic many-body error associated with the CCSD approximation and from the Gaussian basis,
a dominant source of uncertainty comes from the finite size effects due to finite sampling of the Brillouin zone.
This is most evident in the Coulomb matrix elements due to the slowly decaying tail of the Coulomb operator for $r \rightarrow \infty$, $G=0$.
Several corrections for the tail error\cite{gygi_self-consistent_1986, paier_perdewburkeernzerhof_2005, gruber_applying_2018} have been developed to reduce it.
We employed the extrapolation of observable values as a robust and straightforward way to correct the $G=0$ error.
For this, we performed several calculations with different samplings (numbers of $k$-points) of the 2D BZ.
To extrapolate the size of the band gap at the high-symmetry K point $\Delta_\mathrm{K}$ (Fig.~\ref{fig:1}) we performed simulations with up to $N_k^2 = 7 \times 7 = 49$ $k$-points and fitted the data to linear and quadratic polynomials in $1/N_k$.
The corresponding uniform $k$-point grids were chosen to always include the K point explicitly.

The calculated values of $\Delta_\mathrm{K}$ for monolayer \mos2 are presented in Fig.~\ref{fig:1} for both Hartree-Fock (the single-particle energy difference) and CCSD (the difference between EA and IP eigenvalues) as a function of the number of $k$-points along one reciprocal axis $N_k$. 
Two datasets are presented in each case, ``2D'' and ``3D'', standing for two possible approaches to calculating matrix elements $h$ and $v$.
The ``3D'' approach is the conventional way to describe a 2D crystal with PBC in 3 dimensions: periodic images of a 2D material are separated with a large  but finite vacuum in the \textit{z} direction while the integral Eq.~\ref{eq:v} is carried out in 3 dimensions with a 3D Coulomb kernel $K = 4 \pi / G^2$.
The ``2D'' approach assumes PBC in two dimensions only ($xy$ plane) while the $z$ direction is under ``infinite'' boundary conditions and is integrated out in real space. The Coulomb kernel in this case can be found e.g. in Ref.~\cite{sundararaman2013regularization}.

The two approaches are equivalent only in the $N_k \rightarrow \infty$ limit while finite-size errors behave differently in each case.
The finite-size error in the 2D case can be estimated as the magnitude of the first non-zero term in Eq.~\ref{eq:v}:
\begin{equation}
  \mathrm{err}_{2D}(\Delta) \sim w_2 / G_\mathrm{min} \sim 1 / N_k ~,
    \label{eq:e2}
\end{equation}
where $G_\mathrm{min}$ is the spacing of the 2D reciprocal grid and $w_2$ is the surface area of the reciprocal 2D grid.
The same approach in 3D yields a different error estimate:
\begin{equation}
  \mathrm{err}_{3D}(\Delta) \sim w_3 / G_\mathrm{min}^2 \sim \max (N_k, z)^2 / (z N_k^2) ~,
    \label{eq:e3}
\end{equation}
where $w_3$ is the volume of the reciprocal 3D box and $z$ stands for the ratio between the vacuum size and the lattice constant\footnote{Structural parameters are taken from Ref.~\cite{zhu_giant_2011}}: $z \sim 20 \mathrm{\AA} / 3.193 \mathrm{\AA} \approx 6.26$ in the \mos2 model presented.
Eq.~\ref{eq:e3} takes two possible limits: $\mathrm{err}_{3D}(\Delta) \sim z / N_k^2$, $z \gg N_k$ and $\mathrm{err}_{3D}(\Delta) \sim 1 / z$, $z \ll N_k$.
This is consistent with the behavior presented in Fig.~\ref{fig:1}: while the 2D treatment results in a clear $1/N_k$ trend for both Hartree-Fock and CCSD band gap, $\Delta (N_k)$  calculated with
the 3D treatment exhibits a minimum corresponding to the transition from the $z \gg N_k$ to the $z \ll N_k$ limit.
Because of the high computational cost of the CCSD calculations, we were restricted to a maximum $N_{k, \mathrm{max}} = 7$, comparable to the vacuum size $z \sim 6$, prohibiting a meaningful extrapolation of the CCSD band gap in the ``3D'' case.
However, in the case of the HF gap, where we computed up to a larger $N_{k, \mathrm{max}} = 10$, the two approaches agree for the extrapolated HF band gap size (Fig.~\ref{fig:1} (c)). Given the greater reliability of the 2D extrapolation for CCSD with the $k$ point meshes we could sample in this
work, we will henceforth only discuss results from the 2D treatment.

Previous studies have found a range of values for the 2D \mos2 fundamental band gap:
$1.9$ -- $2.2$ eV (STS experiments\cite{vancso_intrinsic_2016, hill_band_2016}),
$1.7$ -- $1.8$ eV (DFT calculations\cite{li_electronic_2007, zhu_giant_2011, johari_tuning_2012}),
$1.8$ -- $2.3$ eV (hybrid functional DFT calculations\cite{ellis_indirect_2011, liu_sulfur_2013, zahid_generic_2013}),
$2.5$ -- $2.8$ eV (DFT-based $GW$ calculations\cite{qiu_optical_2013, qiu_screening_2016}), while the optical band gap ranged from
$1.8$ -- $1.9$ eV (optical spectroscopy\cite{mak_atomically_2010, splendiani_emerging_2010, mak_control_2012, mak_tightly_2013}), and
$1.8$ -- $2.2$ eV ($GW$-BSE calculations\cite{qiu_optical_2013, qiu_screening_2016}).
The extrapolated CCSD band gap value of  $3.00 \pm 0.05$ eV
is close to the fundamental band gap predicted by DFT-$GW$ from Ref.~\cite{ramasubramaniam_large_2012}.
However, CCSD consistently predicts a higher band gap size for other materials in the same family: \mose2, \ws2, \wse2, see Table~\ref{tab:1}.
This is likely related to the absence of spin-orbit splitting of the valence bands in our model.
An estimate of the band gap size $\Delta_\mathrm{K}$ for the spin-orbit splitting is presented in the last row of Table~\ref{tab:1}, where the CCSD band gap value is combined with the $GW$ spin-orbit splitting size.
The corrected band gap sizes of monolayer \mos2, \mose2 and \ws2 are then in
closer agreement with the $GW$ values in Ref.~\cite{ramasubramaniam_large_2012}
while the \wse2 CCSD band gap is $300$ meV larger.

\begin{table}
    \begin{tabularx}{\columnwidth}{c |*{4}{Y}}
        & \mos2        & \mose2 & \ws2  & \wse2 \\
        \hline
        DFT-PBE & 1.60         & 1.35   & 1.56  & 1.19 \\
        $GW$    & 2.82         & 2.41   & 2.88  & 2.42 \\
        CCSD    & 3.00 (2.93)  & 2.63   & 3.23  & 3.01 \\
        CCSD*   & 2.92         & 2.52   & 3.00  & 2.76
    \end{tabularx}
    \caption{
        Band gap sizes across four monolayer TMDs.
        The reference DFT and $GW$ data is taken from Ref.~\cite{ramasubramaniam_large_2012} (note slightly different lattice parameters).
        The value in brackets indicates an indirect $\Gamma$-K band gap. \newline
        *The last row is calculated as $\Delta_\mathrm{CCSD*} = \Delta_\mathrm{CCSD} - \mathrm{SOC}/2$, where $\Delta_\mathrm{CCSD}$ is the extrapolated CCSD band gap at $K$ and $\mathrm{SOC}$ are spin-orbit-coupling-induced valence band splittings taken from Ref.~\cite{ramasubramaniam_large_2012}.
    }
    \label{tab:1}
\end{table}

\begin{figure}
    \includegraphics{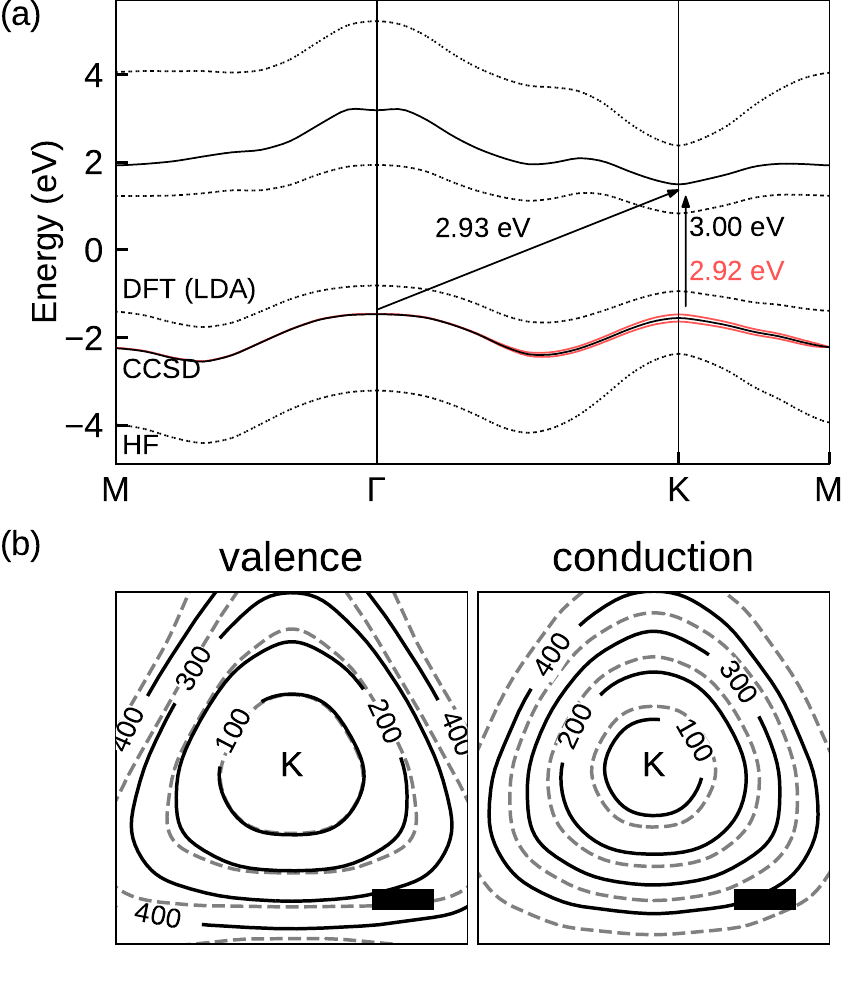}
    \caption{
        Electronic band structure of monolayer \mos2.
        (a) Electronic band structure of monolayer \mos2 along the high-symmetry M-$\Gamma$-K-M path: CCSD bands (solid lines), HF and LDA DFT bands (dashed lines), and the spin-orbit correction to CCSD valence bands (red).
        Arrows indicate CCSD direct and indirect band gaps.
        All calculations were performed within the same setup (lattice parameters, basis sets, all-electron calculation).
        HF and CCSD bands are shifted to match the extrapolated band gap size.
        (b) Shapes of valence and conduction bands at the $\mathrm{K}$ point: CCSD (solid) and LDA DFT (dashed) isolines.
        The numbers indicate isoline level offsets in meV from the corresponding valence and conduction band extrema at the $\mathrm{K}$ point.
        The scale bar is $0.1$ $\mathrm{\AA}^{-1}$.
    }
    \label{fig:2}
\end{figure}

Most earlier theoretical and experimental studies of monolayer \mos2 agree that the fundamental band gap is direct and located at the K, K' points of the hexagonal BZ.
However, bilayer, multi-layer, bulk\cite{kumar_electronic_2012, wang_electronics_2012, cheiwchanchamnangij_quasiparticle_2012, peelaers_effects_2012},
as well as deformed monolayer crystals of \mos2\cite{johari_tuning_2012, peelaers_effects_2012, conley_bandgap_2013, he_experimental_2013}
are also known to have an indirect band gap originating from the the maximum of the valence bands located at the $\Gamma$ point.
In addition, some $GW$ studies\cite{huser_how_2013, shi_quasiparticle_2013} have suggested a different location of the conduction bands minimum occurring along the $\Gamma$-K high-symmetry path but not at the $K$ point itself.
These observations are explained by a rather small $\sim 100$ meV difference between the corresponding band extrema\cite{mak_atomically_2010}.
To investigate this question in detail, we plot the EOM-CCSD electronic band structure of monolayer \mos2 and compare it to other methods in Fig.~\ref{fig:2}.
To obtain the band energies, we performed CCSD band structure calculations using multiple $k$ point grids as described previously in the text.
From these calculations, we obtained the dispersion of spin-degenerate valence and conduction bands adjacent to the band gap with a high ($>90\%$) weight of single-particle excitations. 
The band structure we obtain contains an indirect $\Gamma-\mathrm{K}$ band gap as shown in Fig.~\ref{fig:2}(a).
However, the difference between the $\Gamma-\mathrm{K}$ and $\mathrm{K}-\mathrm{K}$ transitions is $\sim 70$ meV: this is less than a typical spin-orbit-induced valence bands shifts $150 / 2 = 75$ meV\cite{mak_control_2012}.
Thus, a direct band gap $\Delta_\mathrm{K}$ is expected upon the inclusion of the spin-orbit coupling into the model.
Red bands in Fig.~\ref{fig:2} (a) demonstrate the spin-orbit splitting of 164 meV\cite{ramasubramaniam_large_2012} added to the CCSD bands as an illustration of the expected behavior.
For comparison, local-density-approximation (LDA) DFT bands without spin-orbit splitting effects plotted in Fig.~\ref{fig:2}(b) also possess an indirect $1.61$ eV band gap with a $110$ meV difference between valence band extrema at K and $\Gamma$.
In contrast, HF bands form a pronounced direct band gap at the K point, Fig.~\ref{fig:2}(b).
Note that these observations are subject to the choice of lattice parameters: a relatively small strain is known to cause a pronounced indirect band gap in both DFT and $GW$ models of the material\cite{kumar_electronic_2012, shi_quasiparticle_2013} and the same effect is expected here.

As a further analysis of the band structure in Fig.~\ref{fig:2}, we computed the effective electron $m_e^*$ and hole $m_h^*$ masses.
Previous studies have found effective masses to be in a wide energy region depending on the approach\cite{
    ramasubramaniam_large_2012,
    cheiwchanchamnangij_quasiparticle_2012,
    shi_quasiparticle_2013,
    xi_tunable_2014,
    chang_ballistic_2014,
    wickramaratne_electronic_2014}: $m_e^* = 0.3-0.6 m_e$, $\left | m_h^* \right | = 0.35-0.65 m_e$ (experimental value $\left | m_h^* \right | = 0.43-0.48 m_e$\cite{jin_substrate_2015}).
The electron mass $m_e^* = 0.48 m_e$ predicted by CCSD falls within the above range.
In contrast, the hole mass $\left | m_h^* \right | = 0.75 m_e$ deviates significantly due to a less dispersive valence band.
This is partially explained by the aforementioned absence of the spin-orbit splitting in the model, which would otherwise increase the energy scale and reduce the effective mass of the hole quasiparticles.
To support this explanation, we calculated the effective hole mass from the LDA DFT band structure under the spin-restricted approximation and observed a similar effect: $\left | m_h^{*\mathrm{(DFT)}} \right | = 0.70 m_e$.
The similarity of the LDA DFT and CCSD spin-restricted valence bands is evident from Fig.~\ref{fig:2} (b) where the band profiles are plotted: while the effective masses match closely, the trigonal warping of the CCSD valence bands is less pronounced.
Conversely, the CCSD conduction bands are considerably more dispersive while retaining the LDA DFT conduction bands' shape.

\begin{figure}
    \includegraphics{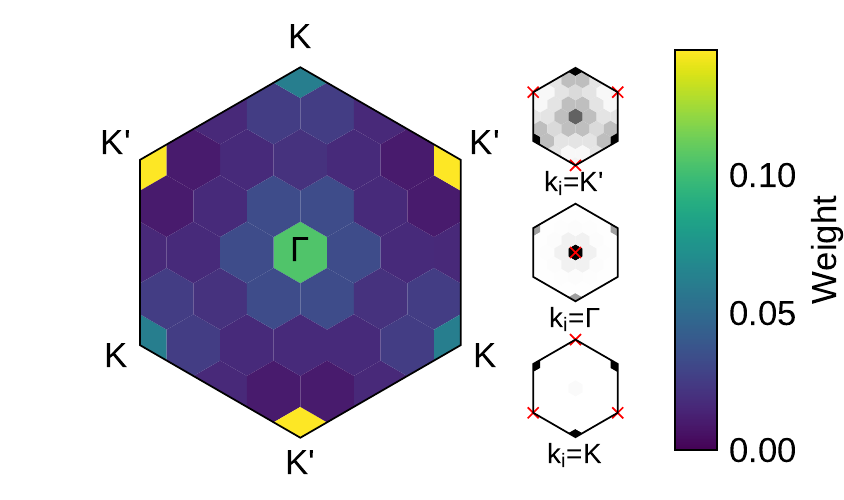}
    \caption{
        Weights of the trion-like IP root (color) in the EOM CCSD spectrum of monolayer \mos2 (two-particle weight $w > 0.99$).
        The central image shows weights of the trion in the hexagonal Brillouin zone (6x6 sampling) $w \left ( k_i \right ) = \sum\limits_{k_j; ija} r^{(2)}_{ija} \left (k_i, k_j \right )$, where $r^{(2)}$ are the 3-particle amplitudes (h+h+e) of the solution.
        The three side images resolve $w \left ( k_j \right ) = \sum\limits_{ija} r^{(2)}_{ija} \left (k_i, k_j \right )$ for the $k_i$ contributing the largest weights in the first plot.
    }
    \label{fig:trion}
\end{figure}

\begin{figure}
    \includegraphics{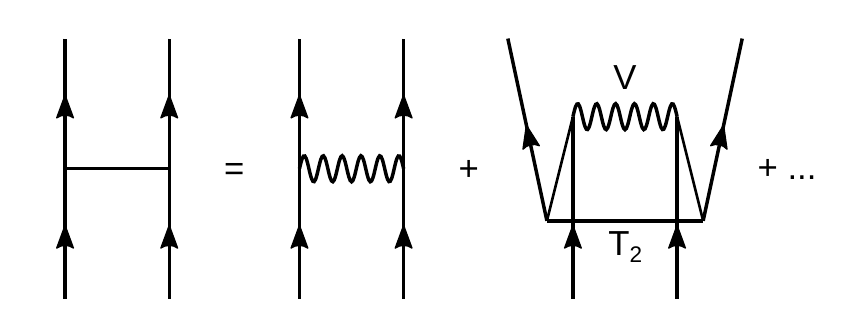}
    \caption{
        Diagrammatic expansion of the effective particle-particle interaction in EOM-CCSD.
        The terms shown in the Figure are bare Coulomb interaction (middle) and one of the $T_2$ contractions from the CCSD theory (right).
    }
    \label{fig:diagram}
\end{figure}

Finally, we examine the EOM-CCSD excitations that display an explicit many-body nature.
Trions in monolayer \mos2 and similar materials have been observed in several experiments\cite{mak_tightly_2013, ross_electrical_2013, zhang_absorption_2014, christopher_long_2017, gaur_manipulation_2019, li_direct_2019} where a large ten-meV-order binding energy between charged and charge-neutral excitations was measured.
We focused on trion roots in our model that are closest to the Fermi level (example shown in Fig.~\ref{fig:trion}).
All trions we observed couple single-particle states belonging to the topmost valence band and lowest-lying conduction band.
The primary contributions to the EOM amplitudes originate from either K and K' points (both electrons and holes) or the $\Gamma$ point (holes only) which are closest to the Fermi level in our model, as expected.
Fig.~\ref{fig:trion} demonstrates an example of such a root where the primary contribution comes from an exciton at the K' point which is coupled to a hole at the K point.
In addition, there is a significant contribution of a double-hole excitation at $\Gamma$ plus an electron at K' to this root.
The trion solution shown in Fig.~\ref{fig:trion} is nearly degenerate with at least 5 other similar roots with total momentum $k=\mathrm{K},\mathrm{K'}$ pointing to the existence of dark excitons in the material\cite{li_direct_2019}.
However, the energies of the trions roughly correspond to those in the
Hartree-Fock single-particle picture $\epsilon_{i} + \epsilon_{j} - \epsilon_{a}$.
This results from the inability of EOM-CCSD to describe screening in three-particle excitations. To understand this, note that
while including the pph and hhp spaces when diagonalizing the EOM-CC effective Hamiltonian results in effective
screening of the principal poles of the Green's function~\cite{lange2018relation}, the effective Hamiltonian
matrix elements do not themselves properly contain screened interactions. For example, all Goldstone diagrams
contributing to the particle-particle effective interactions always contain at least one unscreened interaction line, see Fig.~\ref{fig:diagram} for an example.
To screen the particle/hole interactions in the trion thus requires including the 3p2h and 3h2p spaces
when diagonalizing the EOM-CC effective Hamiltonian, which incurs greater expense.

In summary,  we have found that the equation-of-motion coupled cluster within the singles and doubles (EOM-CCSD) approximation
can be applied in practice to access the electronic
band gap and band structure of monolayer TMDs.
We analyzed the form of the convergence to the thermodynamic limit in two dimensions, which is the primary source of error, similarly to seen in 3D semiconductors\cite{mcclain_gaussian-based_2017}.
The size of the band gaps are close to those in previous $GW$ studies of 2D TMDs with a slight trend of increasing the band gap size.
The discrepancies between our computed electronic band structure for monolayer \mos2 and experimental and earlier theoretical results can be
traced to the neglect of spin-orbit coupling in our Hamiltonian, known to be important for this system.
The CCSD approach is able to qualitatively predict three-particle excited states, although quantitative energies will require the
inclusion of more states when diagonalizing the effective Hamiltonian.
Our results illustrate the potential for many-body coupled cluster methods to provide a deeper understanding of the
electronic structure of novel two-dimensional materials.

\textit{Acknowledgments.}
Author A.P. thanks Swiss NSF for the support provided through the Early Postdoc. Mobility program (project P2ELP2\_175281).
Author GKC was supported by the US Department of Energy via SC0018140 for this work.
Secondary support for PySCF software infrastructure was provided by the US National Science Foundation under NSF SI2-1657286.
We thank Q. Sun and J. McClain for valuable discussions.

\bibliography{main}

\end{document}